\documentclass[11pt]{article}

\def\mytitle#1{\setcounter{equation}{0}
\setcounter{footnote}{0}
\begin{flushleft}\Large\textbf{#1}\end{flushleft}
\vspace{0.25cm}}
\def\myname#1{\leftline{{\large #1}}\vspace{-0.13cm}}
\def\myplace#1#2{\small\begin{flushleft}\textit{#1}\\
\texttt{#2}\end{flushleft}}

\usepackage{graphicx}% Include figure files
\begin{document}

\mytitle{Towards a possible solution for the coincidence problem:
\textit{f(G)} gravity as background}

\vskip0.2cm \myname{Prabir Rudra\footnote{prudra.math@gmail.com}}
\myplace{Department of Mathematics, Indian Institute of
Engineering Science and Technology, Shibpur, Howrah-711 103, India.\\
Department of Mathematics, Pailan College of Management and
Technology, Bengal Pailan Park, Kolkata-700 104, India.}{}

\begin{abstract}
In this article we address the well-known cosmic coincidence
problem in the framework of the \textit{f(G)} gravity. In order to
achieve this, an interaction between dark energy and dark matter
is considered. A set-up is designed and a constraint equation is
obtained which generates the \textit{f(G)} models that do not
suffer from the coincidence problem. Due to the absence of a
universally accepted interaction term introduced by a fundamental
theory, the study is conducted over three different forms of
logically chosen interaction terms. To illustrate the set-up three
widely known models of \textit{f(G)} gravity are taken into
consideration and the problem is studied under the designed
set-up. The study reveals that the popular $f(G)$ gravity models
does not approve of a satisfactory solution of the long standing
coincidence problem, thus proving to be a major setback for them
as successful models of universe. Finally, two non-conventional
models of \textit{f(G)} gravity have been proposed and studied in
the framework of the designed set-up. It is seen that a complete
solution of the coincidence problem is achieved for these models.
The study also reveals that the b-interaction term is much more
preferable compared to the other interactions, due to its greater
compliance with the recent observational data.
\end{abstract}

\vspace{5mm}


\vspace{5mm}

{\it Pacs. No.: 95.36.+x, 95.35.+d}\\

\vspace{5mm}

\section{Introduction}
Recent observational evidences from Ia supernovae, CMBR via WMAP,
galaxy redshift surveys via SDSS indicated that the universe is
going through an accelerated expansion of late \cite{Perlmutter1,
Spergel1, Bennett1, Tegmark1, Allen1}. With this discovery the
incompatibility of general relativity (GR) as a self sufficient
theory of gravity came into light. Since no possible explanation
of this phenomenon could be attributed in the framework of
Einstein's GR, a proper modification of the theory was required
that will successfully incorporate the late cosmic acceleration.
As the quest began, two different approaches regarding this
modification came into existence.

According to the first approach, cosmic acceleration can be
phenomenally attributed to the presence of a mysterious negative
energy component popularly known as \textit{dark energy (DE)}
\cite{Riess1}. Here we modify the right hand side of the
Einstein's equation, i.e. in the matter sector of the universe.
Latest observational data shows that the contribution of DE to the
energy sector of the universe is $\Omega_{d}=0.7$. With the
passage of time, extensive search saw various candidates for DE
appear in the scene. Some of the popular ones worth mentioning are
Chaplygin gas models \cite{Kamenschik1, Gorini1}, Quintessence
Scalar field \cite{Ratra1}, Phantom energy field \cite{Caldwell1},
etc. A basic feature of these models is that, they violate the
strong energy condition i.e., $\rho+3p<0$, thus producing the
observed cosmic acceleration. Recent reviews on DE can be found in
\cite{Joyce1, Bam1}

A different section of cosmologists resorted to an alternative
approach for explaining the expansion. This concept is based on
the modification of the gravity sector of GR, thus giving birth to
modified gravity theories. A universe associated with a tiny
cosmological constant, i.e. the $\Lambda$CDM model served as a
prototype for this approach. It was seen that the model could
satisfactorily explain the recent cosmic acceleration and passed a
few solar system tests as well. But with detailed diagnosis it was
revealed that the model was paralyzed with a few cosmological
problems. Out of these, two major problems that crippled the model
till date are the Fine tuning problem (FTP) and the Cosmic
Coincidence problem (CCP). The FTP refers to the large discrepancy
between the observed values and the theoretically predicted values
of cosmological parameters. Numerous attempts to solve this
problem can be found in the literature. Among them, the most
impressive attempt was undertaken by Weinberg in \cite{Weinberg1}.
Although the approaches for the solutions are different, yet,
almost all of them are basically based on the fact that the
cosmological constant may not assume an extremely small static
value at all times during the evolution of the universe (as
predicted by GR), but its nature should be rather dynamical
\cite{Bisabr1}. These drawbacks reduced the effectiveness of the
model, as well as its acceptability, and hence alternative
modifications of gravity was sought for. Some of the popular
models of modified gravity that came into existence in recent
times are loop quantum gravity \cite{Rovelli1, Ashtekar1}, Brane
gravity \cite{Brax1, Maartens1, Maartens2}, \textit{f(R)} gravity
\cite{Kerner1, Allemandi1, Carroll1}, \textit{f(T)} gravity
\cite{Linder1, Li1, Miao1, Li2}, etc. Reviews on extended gravity
theories can be found in \cite{Capo1,Noji1}.

In this work we will consider \textit{f(G)} model as the theory of
gravity \cite{Noj1, Cognola1}. Over the years, several
modifications to GR have been achieved, by generalizing the
Einstein-Hilbert Lagrangian used in GR. $F(R)$ and $F(T)$
gravities are common examples of such modifications. Gauss-Bonnet
(GB) modification to GR is another way of modifying the Einstein
gravity that has gained popularity over the past few years,
because it is considered as the low energy limit of string theory.
In this modification, one generally adds quadratic terms,
specifically the GB terms, which involve second order curvature
invariants. But as it turns out to be, the GB term becomes trivial
in 4-dimension, and hence, it is used from another form of
modified gravity, namely, the modified GB gravity \cite{Noj1}.
Here an arbitrary function of the GB term, $f(G)$ is added to the
Einstein-Hilbert Lagrangian, to bring about the modification.

In \cite{Bisabr2} Bisabr studied cosmological coincidence problem
in the background of \textit{f(R)} gravity. Motivated by Bisabr's
work, we dedicate the present assignment to the study of the
coincidence problem in \textit{f(G)} gravity. The paper is
organized as follows: Basic equations of \textit{f{G}} gravity are
furnished in section 2. In section 3, we discuss the coincidence
problem. The set-up for the present study is discussed in section
4. We illustrate the designed set-up by a few examples in section
5, and finally the paper ends with a short conclusion in section
6.

\section{Basic equations of \textit{f(G)} gravity}
The 4-dimensional action in \textit{f(G)} gravity is  given by,
\begin{equation}\label{1}
S=\frac{1}{\kappa^2}\int
\sqrt{-g}\left[\frac{R}{2}+f(G)\right]d^{4}x+S_{m}
\end{equation}
where $R$ is the Ricci scalar curvature, $f(G)$ is a generic
function of the Gauss-Bonnet topological invariant $G$,
$\kappa^{2}=8\pi G$ and $S_{m}$ is the matter action.

Varying the above action with respect to the metric one can obtain
the field equation as,
\begin{equation}\label{2}
G_{\mu\nu}+8\left[R_{\mu\rho\nu\sigma}+R_{\rho\nu}g_{\sigma\mu}+\frac{1}{2}\left(g_{\mu\nu}g_{\sigma\rho}
-g_{\mu\sigma}g_{\nu\rho}\right)R-R_{\rho\sigma}g_{\nu\mu}-R_{\mu\nu}g_{\sigma\rho}+R_{\mu\sigma}g_{\nu\rho}\right]\nabla^{\rho}\nabla^{\sigma}F'(G)+\left(GF'(G)-F\right)g_{\mu\nu}=T_{\mu\nu}^{m}
\end{equation}
where $G_{\mu\nu}$ is the Einstein tensor, $T_{\mu\nu}^{m}$ is the
energy momentum tensor of matter. Here we consider
$\kappa^{2}=8\pi G=1$ and prime denotes ordinary derivative with
respect to $G$. For spatially flat Robertson-Walker metric
\begin{equation}\label{3}
ds^{2}=-dt^{2}+a(t)^{2}dx^{2}
\end{equation}
we have
\begin{equation}\label{4}
R=6\left(\dot{H}+2H^{2}\right),
~~~~~~~~~~~~~G=24H^{2}\left(\dot{H}+H^{2}\right)
\end{equation}
where $H$ is the Hubble parameter and dot denotes the time
derivative. Considering the universe to be filled with
pressureless dark matter under the assumption of a flat universe
the modified Friedmann equations for \textit{f(G)} gravity are,

\begin{equation}\label{5}
3H^{2}=GF'(G)-F-24H^{3}\dot{F}'(G)+\rho_{m}
\end{equation}
\begin{equation}\label{6}
-2\dot{H}=-8H^{3}\dot{F}'(G)+16H\dot{H}\dot{F}'(G)+8H^{2}\ddot{F}'(G)+\rho_{m}
\end{equation}

The energy conservation equations are given by,
\begin{equation}\label{7}
\dot{\rho_{m}}+3H\rho_{m}=Q
\end{equation}
\begin{equation}\label{8}
\dot{\rho_{G}}+3H\left(1+\omega_{G}\right)\rho_{G}=-Q
\end{equation}

Here $\omega_{G}=\frac{p_{G}}{\rho_{G}}$ is the EoS parameter of
the energy sector and $Q$ is the interaction between the matter
and the energy sector of the universe. The EoS parameter is given
by,

\begin{equation}\label{9}
\omega_{G}=-1+\frac{\left(8H^{2}\ddot{G}+16H\dot{H}\dot{G}-8H^{3}\dot{G}\right)F''+8H^{2}\dot{G}^{2}F'''}{GF'-F-24H^{3}\dot{G}F''}
\end{equation}

\section{The Coincidence problem}

The cosmic coincidence problem has been a serious issue in recent
times regarding various dark energy models. Recent cosmological
observations have indicated that the densities of the matter
sector and the DE sector of the universe are almost identical in
late times. It is known that the matter and the energy component
of the universe have evolved independently from different mass
scales in the early universe, then how come they reconcile to
identical mass scales in the late universe! This is major problem
having its roots in the very formation of the theory. Almost all
the DE models known till date more or less suffer from this
phenomenon.

Various attempts to solve the coincidence problem can be found in
literature. Among them the most impressive are the ones which use
the concept of a suitable interaction between the matter and the
dark energy components of the universe, as given in the
conservation equations (\ref{7}) and (\ref{8}). In this approach,
it is considered that the two sectors of the universe have not
evolved independently from different mass scales. Instead they
evolve together, interacting with each other, allowing a mutual
flow of matter and energy between the two components. Due to this
exchange, the densities of the two components coincide in the
present universe. Although the concept seems to be a really
promising one, yet a problem persists. There is no universally
accepted interaction term, introduced by a fundamental theory
known till date.

It is known that both dark energy and dark matter are not
universally accepted facts, but concepts which are still at the
speculation level. Due to this unknown nature of both dark energy
and dark matter, it is not possible to derive an expression for
the interaction term ($Q$) from the first principles. Such a
situation, demands us to use our logical reasoning and propose
various expressions for $Q$ that will be reasonably acceptable.
The late time dominating nature of dark energy indicates that $Q$
must be considered a small and positive value. On the other hand a
large negative value of interaction will make the universe dark
energy dominated from the early times, thus leaving no scope for
the condensation of galaxies. So the most logical choice for
interaction should contain a product of energy density and the
hubble parameter, because it is not only physically but also
dimensionally justified. So $Q=Q(H\rho_{m}, H\rho_{de})$, where
$\rho_{de}$ is the dark energy density. Since here we are not
planning to add any dark energy by hand, so the effective density
resulting from the extra terms of the modified GB gravity,
$\rho_{G}$ will replace $\rho_{de}$. This leads us to three basic
forms of interactions as given below \cite{del Campo1}:
\begin{equation}\label{10}
b-model:~ Q=3b H\rho_{m}~~~~~~~~~\eta-model:~ Q=3\eta
H\rho_{G}~~~~~~~~\Gamma-model:~ Q=3\Gamma
H\left(\rho_{m}+\rho_{G}\right),~~~~~~~~
\end{equation}
where $b$, $\eta$ and $\Gamma$ are the coupling parameters of the
respective interaction models.

It is worth mentioning that due to its simplicity the most widely
used interaction model is the $b$-model and is available widely in
literature \cite{Berger1, del Campo1, Rudra1, Jamil1}.

\section{The set-up}
In this note we address the coincidence problem in \textit{f(G)}
gravity. \textit{f(G)} gravity has evolved over the past decade as
a candidate for modified gravity theory. From the literature it is
known that \textit{f(G)} gravity is itself self competent in
producing the late cosmic acceleration without resorting to any
forms of dark energy. Therefore in order to keep it simple and
reasonable, we do not consider any separate dark energy components
in the present study. The extra terms of the modified GB gravity
provides the exotic nature and is considered as the dark energy.
We consider the ratio of the densities of matter and dark energy
as, $r\equiv \rho_{m}/\rho_{G}$. Our aim is to devise a set-up
that will aim towards a possible solution to the coincidence
problem. We also want to set up a filtering process that will
separate the favorable $f(G)$ models, that produce a stationary
value of the ratio of the component densities, $r$ from the
unfavorable ones that do not. The time evolution of $r$ is as
follows,
\begin{equation}\label{11}
\dot{r}=\frac{\dot{\rho_{m}}}{\rho_{G}}-r\frac{\dot{\rho_{G}}}{\rho_{G}}
\end{equation}
Using eqns. (\ref{7}), (\ref{8}) and (\ref{11}), we obtain
\begin{equation}\label{12}
\dot{r}=3Hr\omega_{G}+\frac{Q}{\rho_{G}}\left(1+r\right)
\end{equation}

Using the b-interaction given in eqn.(\ref{10}), we get the
expression for $\dot{r}$ as,
\begin{equation}\label{13}
\dot{r}=3Hr\left(b+br+\omega_{G}\right)
\end{equation}
where $\omega_{G}$ is given by eqn.(\ref{9}). Now in order to
comply with observations, it is required that universe should
approach a stationary stage, where either $r$ becomes a constant
or evolves slower than the scale factor. In order to satisfy this
$\dot{r}=0$ in the present epoch, it leads to the following
equation,
\begin{equation}\label{14}
g_{1}(f,H,r_{s},q)=0
\end{equation}
where

$$g_{1}(f,H,r_{s},q)={3Hr_{s}}
\left[b+\frac{{\dot{H}}}{H^2}+q+br_{s}+\frac{1}{{\left(-f[G]+G
f'[G]-576 H^3 \left(\ddot{H} H^2+2 \dot{H} \left(\dot{H} H+2
H^3\right)\right) f''[G]\right)}}\times\right.$$

$$\left.\left(\left(384{\dot{H}}H \left({\ddot{H}} H^2+2 {\dot{H}}
\left({\dot{H}}H+2H^3\right)\right)-192H^3 \left({\ddot{H}} H^2+2
{\dot{H}}\left({\dot{H}} H+2 H^3\right)\right)+192 H^2
\left(\stackrel{...}H H^2+2 {\dot{H}}^2
\left({\dot{H}}\right.\right.\right.\right.\right.$$
\begin{equation}\label{15}
\left.\left.\left.\left.\left.+6 H^2\right)+2 {\ddot{H}} \left(3
{\dot{H}} H+2 H^3\right)\right)\right)f''[G]+4608
H^2\left({\ddot{H}} H^2+2\dot{H} \left(\dot{H} H+2
H^3\right)\right)^2 f'''[G]\right)\right]
\end{equation}
and $r_{s}$ is the value of $r$ when it takes a stationary value.

Using the $\eta$-interaction given in eqn.(10), we get the
expression for $\dot{r}$ as,
\begin{equation}\label{16}
\dot{r}=3H\left[\eta+r\left(\eta+\omega_{G}\right)\right]
\end{equation}
where $\omega_{G}$ is given by eqn.(9). In order to satisfy this
$\dot{r}=0$ in the present epoch, it leads to the following
equation,
\begin{equation}\label{17}
g_{2}(f,H,r_{s},q)=0
\end{equation}
where

$$g_{2}(f,H,r_{s},q)=3H \left[\eta+r_{s}
\left(\frac{\dot{H}}{H^2}+q+\eta+\frac{1}{\left(-f[G]+G f'[G]-576
H^3 \left(\ddot{H} H^2+2 \dot{H} \left(\dot{H} H+2
H^3\right)\right) f''[G]\right)}\times\right.\right.$$

$$\left.\left.\left(\left(384\dot{H} H
\left(\ddot{H} H^2+2 \dot{H} \left(\dot{H} H+2
H^3\right)\right)-192 H^3 \left(\ddot{H} H^2+2 \dot{H}
\left(\dot{H} H+2 H^3\right)\right)+ 192 H^2 \left(\stackrel{...}H
H^2+2 \dot{H}^2 \left(\dot{H}+6
H^2\right)\right.\right.\right.\right.\right.$$

\begin{equation}\label{18}
\left.\left.\left.\left.\left.+2 \ddot{H} \left(3\dot{H} H+2
H^3\right)\right)\right) f''[G]+4608 H^2 \left(\ddot{H} H^2+2
\dot{H} \left(\dot{H} H+2 H^3\right)\right)^2
f'''[G]\right)\right)\right]
\end{equation}

Using the $\Gamma$-interaction given in eqn.(10), we get the
expression for $\dot{r}$ as,
\begin{equation}\label{19}
\dot{r}=3H\left[\Gamma
r^{2}+r\left(2\Gamma+\omega_{G}\right)+\Gamma\right]
\end{equation}
where $\omega_{G}$ is given by eqn.(9). In this case, in order to
satisfy $\dot{r}=0$ in the present epoch, it leads to the
following equation,
\begin{equation}\label{20}
g_{3}(f,H,r_{s},q)=0
\end{equation}
where

$$g_{3}(f,H,r_{s},q)= 3H\left[\Gamma+r_{s}^2\Gamma+r_{s}
\left(\frac{\dot{H}}{H^2}+q+2\Gamma+ \frac{1}{\left(-f[G]+G
f'[G]-576 H^3 \left(\ddot{H} H^2+2 \dot{H} \left(\dot{H} H+2
H^3\right)\right) f''[G]\right)}\times\right.\right.$$

$$\left.\left.\left(\left(384\dot{H} H \left(\ddot{H} H^2+2 \dot{H}
\left(\dot{H} H+2 H^3\right)\right)- 192 H^3 \left(\ddot{H} H^2+2
\dot{H} \left(\dot{H} H+2 H^3\right)\right)+192H^2
\left(\stackrel{...}H H^2+2\dot{H}^2 \left(\dot{H}+6
H^2\right)\right.\right.\right.\right.\right.$$

\begin{equation}\label{21}
\left.\left.\left.\left.\left.+2 \ddot{H} \left(3 \dot{H} H+2
H^3\right)\right)\right) f''[G]+4608 H^2 \left(\ddot{H}
H^2+2\dot{H} \left(\dot{H} H+2 H^3\right)\right)^2
f'''[G]\right)\right)\right]
\end{equation}
In our analysis we will consider $H_{0}$, $r_{0}$ and $q_{0}$ as
the present day values of $H$, $r$ and $q$ respectively. As far as
$q$ is concerned, we start from the best fit parametrization
obtained directly from observational data. Here we use a two
parameter reconstruction function for $q(z)$ \cite{Gong1, Gong2}
\begin{equation}\label{22}
q(z)=\frac{1}{2}+\frac{q_{1}z+q_{2}}{\left(1+z\right)^{2}}
\end{equation}
On fitting this model to Gold data set, we get
$q_{1}=1.47_{-1.82}^{+1.89}$ and $q_{2}=-1.46\pm 0.43$
\cite{Gong2}. We consider $z_{0}=0.25$ and using these values in
eqn.(\ref{23}), we get $q_{0}\approx -0.2$. From recent
observations, we obtain $r_{0}\equiv
\frac{\rho_{m}(z_{0})}{\rho_{T}(z_{0})}\approx \frac{3}{7}$
\cite{Zlatev1, Wei1, Yang1}. The present value of Hubble
parameter, $H_{0}$ is taken as 72, in accordance with the latest
observational data.

\section{Illustration}

We consider the scale-factor, $a$ as a power-law form of time, $t$
as given below,
\begin{equation}\label{23}
a=a_{0}t^{n}
\end{equation}

In order to illustrate the above set-up we consider three
different $f(G)$ gravity models found widely in literature and
test them for the coincidence phenomenon. The three models used are \cite{Nojiri1, Bamba2, Schmidt1}:\\

\vspace{5mm} ~~~~~~~~~~~~~~\textbf{Model1:}
\begin{equation}\label{24}
F(G)=\alpha G^{m_{1}}+\beta G \ln G
\end{equation}
where $\alpha$, $\beta$ and $m_{1}$ are constants, whose values
depend on the cosmographic parameters \cite{Schmidt1}.

\vspace{5mm}~~~~~~~~~~~~~~\textbf{Model2:}
\begin{equation}\label{25}
F(G)=\frac{\alpha_{1}G^{m_{2}}+b_{1}}{\alpha_{2}G^{m_{2}}+b_{2}}
\end{equation}
where $\alpha_{1}$, $\alpha_{2}$, $m_{2}$, $b_{1}$ and $b_{2}$ are
constants.

\vspace{5mm}~~~~~~~~~~~~~~\textbf{Model3:}
\begin{equation}\label{26}
F(G)=a_{3}G^{m_{3}}\left(1+b_{3}G^{m_{4}}\right)
\end{equation}
where $a_{3}$, $m_{3}$, $m_{4}$ and $b_{3}$ are constants.

Using the model 1, i.e., (\ref{24}) and eqn. (\ref{23}) in eqn.
(\ref{15}), we get the following expression for the dynamical
quantity $g_{1}$,

$$g_{1}^{model1}=\frac{1}{t}3nr
\left[b-\frac{1}{n}+q+br+\left[\left(-24^{m_{1}}
\left(-\frac{n^3}{t^4}+\frac{n^4}{t^4}\right)^{m_{1}} \alpha-
\frac{1}{t^3}576 n^3 \left(\frac{2 n^3}{t^5}-\frac{2 n
\left(-\frac{n^2}{t^3}+\frac{2
n^3}{t^3}\right)}{t^2}\right)\right.\right.\right.$$

$$\left.\left.\left.\left(24^{-2+m_{1}} (-1+m_{1}) m_{1}
\left(-\frac{n^3}{t^4}+\frac{n^4}{t^4}\right)^{-2+m_{1}} \alpha
+\frac{\beta}{24
\left(-\frac{n^3}{t^4}+\frac{n^4}{t^4}\right)}\right)- 24
\left(-\frac{n^3}{t^4}+\frac{n^4}{t^4}\right) \beta Log\left[24
\left(-\frac{n^3}{t^4}+\frac{n^4}{t^4}\right)\right]\right.\right.\right.$$

$$\left.\left.\left.+24
\left(-\frac{n^3}{t^4}+\frac{n^4}{t^4}\right) \left(24^{-1+m_{1}}
m_{1} \left(-\frac{n^3}{t^4}+\frac{n^4}{t^4}\right)^{-1+m_{1}}
\alpha +\beta +\beta  Log\left[24
\left(-\frac{n^3}{t^4}+\frac{n^4}{t^4}\right)\right]\right)\right)\right]^{-1}\times\right.$$

$$\left.\left(\frac{1}{t^2}4608 n^2 \left(\frac{2 n^3}{t^5}-
\frac{2 n \left(-\frac{n^2}{t^3}+\frac{2
n^3}{t^3}\right)}{t^2}\right)^2 \left(24^{-3+m_{1}} (-2+m_{1})
(-1+m_{1}) m_{1}
\left(-\frac{n^3}{t^4}+\frac{n^4}{t^4}\right)^{-3+m_{1}}
\alpha-\frac{\beta}{576
\left(-\frac{n^3}{t^4}+\frac{n^4}{t^4}\right)^2}\right)\right.\right.$$

$$\left.\left.+
\left(-\frac{384 n^2 \left(\frac{2 n^3}{t^5}-\frac{2 n
\left(-\frac{n^2}{t^3}+\frac{2
n^3}{t^3}\right)}{t^2}\right)}{t^3}-\frac{192 n^3 \left(\frac{2
n^3}{t^5}-\frac{2 n \left(-\frac{n^2}{t^3}+\frac{2
n^3}{t^3}\right)}{t^2}\right)}{t^3}+\frac{192 n^2 \left(-\frac{6
n^3}{t^6}+\frac{2 n^2 \left(-\frac{n}{t^2}+\frac{6
n^2}{t^2}\right)}{t^4}+\frac{4 n \left(-\frac{3 n^2}{t^3}+\frac{2
n^3}{t^3}\right)}{t^3}\right)}{t^2}\right)\right.\right.$$
\begin{equation}
\left.\left.\left(24^{-2+m_{1}} (-1+m_{1}) m_{1}
\left(-\frac{n^3}{t^4}+\frac{n^4}{t^4}\right)^{-2+m_{1}}\alpha
+\frac{\beta}{24
\left(-\frac{n^3}{t^4}+\frac{n^4}{t^4}\right)}\right)\right)
\right]
\end{equation}
Similarly expressions for $g_{1}$ is obtained for the other two
$f(G)$ models. Expressions for $g_{2}$ and $g_{3}$ are also found
for all the three gravity models. As it can be seen from above
that the expressions are really lengthy, so we do not include all
of them in the manuscript.

We have generated plots for $g_{1}$, $g_{2}$ and $g_{3}$ against
cosmic time, $t$ for each of the three models in the figures 1, 2
and 3, for all the three forms of interactions, $b$, $\eta$ and
$\Gamma$. Particular numerical values for the involved parameters
have been considered which are in accordance with the recent
observational data \cite{Nojiri1, Bamba2, Schmidt1}. Moreover in
figs 4 and 5, we have illustrated two non-conventional models of
$f(G)$ gravity i.e., the logarithmic model and the exponential
model, which gives very interesting results when used in the
designed set-up.

\vspace{2mm}
\begin{figure}
~~~~~~~~~~~~~~~~~\includegraphics[height=2.5in]{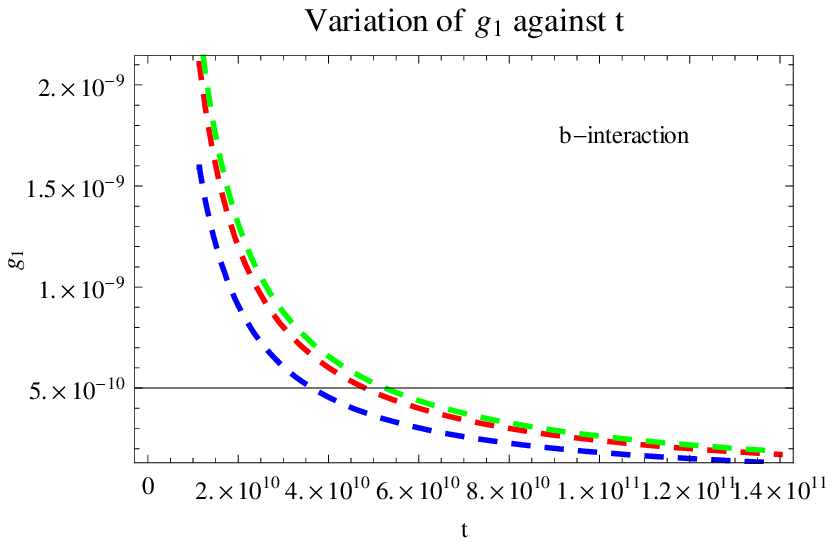}~~~~\\
\vspace{5mm}
~~~~~~~~~~~~~~~~~~~~~~~~~~~~~~~~~~~~~~~~~~~~~~~~~~~~~~~Fig.1~~~~~~~~~~~~~~~~~~~~~~~~~~~~~~~~~\\
\vspace{1mm} \textsl{Fig 1 : The plot of
$g_{1}(f_{0},H_{0},r_{s0},q_{0})$ against time $t$ for model1
(red), model2 (blue) and model3 (green) using $b$ interaction. The
other parameters are considered as $q=-0.2, r=3/7, \alpha=1,
\beta=4,
n=10, m_{1}=0.2, b=1.5, a_{1}=-1, b_{1}=-1, a_{2}=2, b_{2}=0.5, m_{2}=1.5, a_{3}=-1, b_{3}=0, m_{3}=1.5, m_{4}=0.$}\\
\end{figure}

\vspace{3mm}

\begin{figure}
~~~~~~~~~~~~~~~~~\includegraphics[height=2.5in]{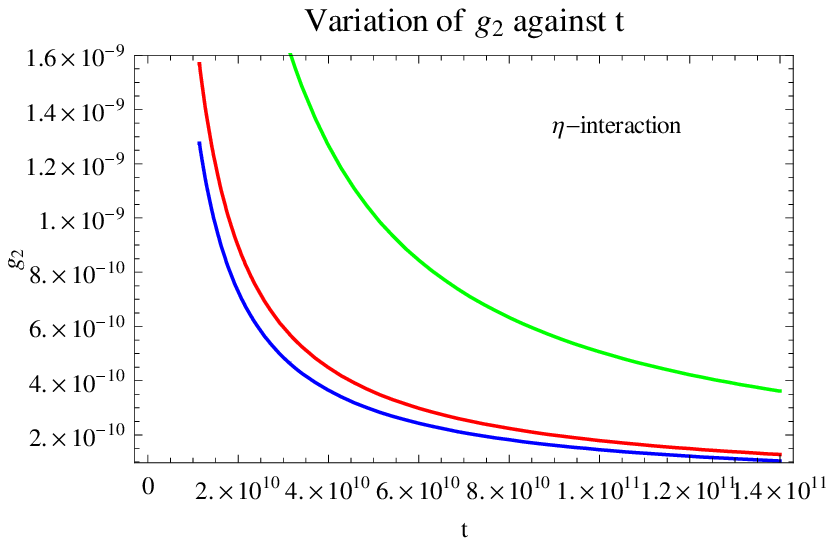}~~~~\\
\vspace{5mm}
~~~~~~~~~~~~~~~~~~~~~~~~~~~~~~~~~~~~~~~~~~~~~~~~~~~~~~~Fig.2~~~~~~~~~~~~~~~~~~~~~~~~~~~~~~~~~\\
\vspace{1mm} \textsl{Fig 2 : The plot of
$g_{2}(f_{0},H_{0},r_{s0},q_{0})$ against time $t$ for model1
(red), model2 (blue) and model3 (green) using $\eta$ interaction.
The other parameters are considered as $q=-0.2, r=3/7, \alpha=1,
\beta=4,
n=10, m_{1}=0.2, \eta=1.5, a_{1}=-1, b_{1}=-1, a_{2}=2, b_{2}=0.5, m_{2}=1.5, a_{3}=-1, b_{3}=0, m_{3}=1.5, m_{4}=0.$}\\
\end{figure}

\vspace{2mm}
\begin{figure}
~~~~~~~~~~~~~~~~~\includegraphics[height=2.5in]{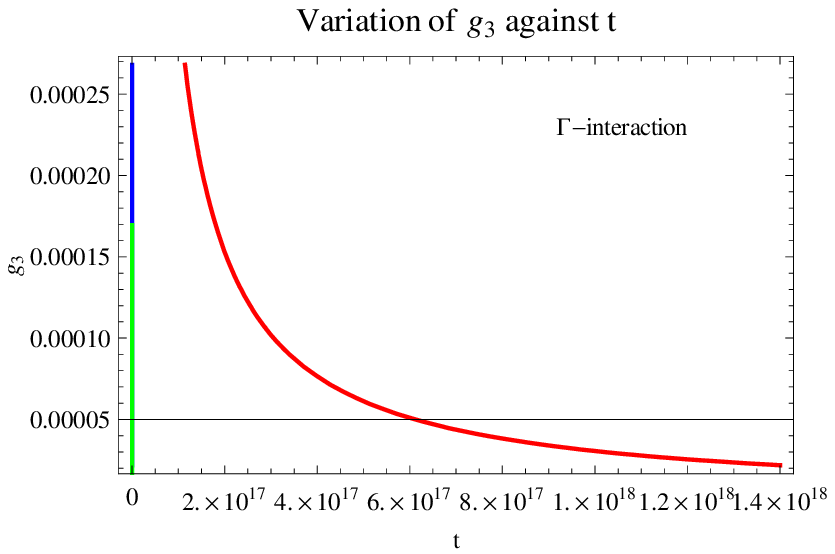}~~~~\\
\vspace{5mm}
~~~~~~~~~~~~~~~~~~~~~~~~~~~~~~~~~~~~~~~~~~~~~~~~~~~~~~~Fig.3~~~~~~~~~~~~~~~~~~~~~~~~~~~~~~~~~\\
\vspace{1mm} \textsl{Fig 3 : The plot of
$g_{3}(f_{0},H_{0},r_{s0},q_{0})$ against time $t$ for model1
(red), model2 (blue) and model3 (green) using $\Gamma$
interaction. The other parameters are considered as $q=-0.2,
r=3/7, \alpha=1, \beta=4,
n=10, m_{1}=0.2, \Gamma=5\times 10^{10}, a_{1}=-1, b_{1}=-1, a_{2}=2, b_{2}=0.5, m_{2}=1.5, a_{3}=-1, b_{3}=0, m_{3}=1.5, m_{4}=0.$}\\
\end{figure}

\vspace{2mm}
\begin{figure}
~~~~~~~~~~~~~~~~~\includegraphics[height=2.5in]{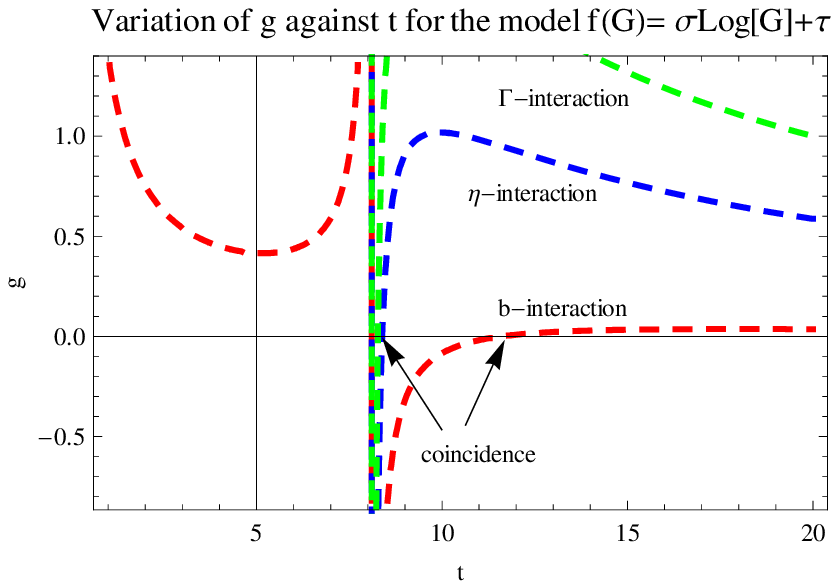}~~~~\\
\vspace{5mm}
~~~~~~~~~~~~~~~~~~~~~~~~~~~~~~~~~~~~~~~~~~~~~~~~~~~~~~~Fig.4~~~~~~~~~~~~~~~~~~~~~~~~~~~~~~~~~\\
\vspace{1mm} \textsl{Fig 4 : The plot of $g$ against time $t$ for
different forms of interaction for the Logarithmic model,
$f(G)=\sigma Log(G)+\tau$. The other parameters are considered as
$q=-0.2, r=3/7, \sigma=5, \tau=0.4,
n=3, b=1.5, \eta=1.5, \Gamma=1.5.$}\\
\end{figure}

\vspace{2mm}
\begin{figure}
~~~~~~~~~~~~~~~~~\includegraphics[height=2.5in]{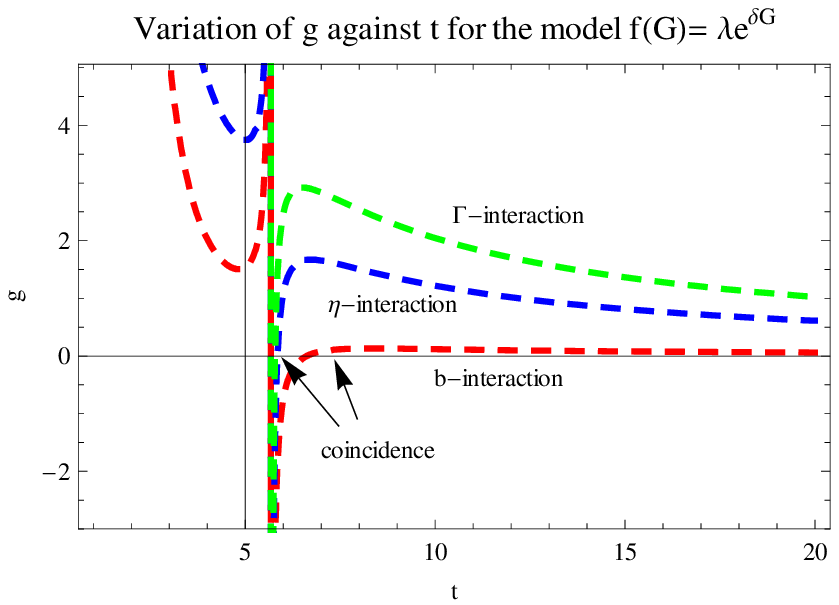}~~~~\\
\vspace{5mm}
~~~~~~~~~~~~~~~~~~~~~~~~~~~~~~~~~~~~~~~~~~~~~~~~~~~~~~~Fig.5~~~~~~~~~~~~~~~~~~~~~~~~~~~~~~~~~\\
\vspace{1mm} \textsl{Fig 5 : The plot of $g$ against time $t$ for
different forms of interaction for the Exponential model,
$f(G)=\lambda \exp(\delta G)$. The other parameters are considered
as $q=-0.2, r=3/7, \lambda=2, \delta=0.4,
n=3, b=1.5, \eta=1.5, \Gamma=1.5.$}\\
\end{figure}

\section{Discussion and conclusion}
From the figures it is evident that the $g$ vs $t$ curves become
asymptotic near the time axis, when the cosmic time corresponds to
the age of the universe, i.e. $14\times 10^{9}$ years. As a result
of this, $g$ never reaches the zero level. Hence $\dot{r}\neq 0$,
makes the realization of a stationary phase extremely difficult.
The asymptotic nature of the curves are indicative of the fact
that as time evolves the trajectories move closer and closer to
the time axis. Therefore for the given models, the coincidence
problem is substantially alleviated with the evolution of time,
but never ever gets solved. This is truly a set back for the
models which are known to satisfy most of the solar system tests.

But from the set-up that we have designed in this assignment, we
can generate as well as filter various models of $f(G)$ gravity
which are completely free from the coincidence problem. Two such
models have been illustrated in the figs.4 and 5. In fig.4, we
have generated the plot of $g$ vs $t$, for the logarithmic model
$(f(G)=\sigma Log(G)+\tau)$ for all the three interaction terms.
It can be seen from the plot that $g$ reaches the zero level for
all the interaction terms at around $t=8$. Particularly for the
b-interaction, the stationary scenario is realized for $t>11$. In
fig.5, a similar plot has been generated for the exponential model
$(f(G)=\lambda \exp(\delta G))$. From the plot, it is evident that
a stationary scenario is achieved at around $t=6$ for all the
interactions. Particularly for the b-interaction a continuous
stationary scenario is realized for $t\geq 7$. So from the above
discussion it is quite clear that for the logarithmic and the
exponential models a complete solution for the coincidence problem
can be achieved following the set-up that we have designed in the
present assignment. Looking at the plots 4 and 5, it must also be
mentioned that the b-interaction term is much more preferable
compared to the other interactions, since it helps us to realize a
continuous stationary phase between dark energy and dark matter
after a certain point of time in the cosmological time-line, thus
complying with
the recent observational data.\\

{\bf Acknowledgement:}\\\\
The author sincerely acknowledges the anonymous referee for his or
her constructive comments which helped the author to improve the
quality of the manuscript.\\

\end{document}